\documentclass[12pt,preprint]{aastex}

\slugcomment{To appear in ApJ, June 2003}
\shorttitle{Spectral Variability of BL Lac Objects}
\shortauthors{Vagnetti et al.}
\begin{document}
\title{Spectral Slope Variability of BL Lac Objects in the Optical Band}
\author{Fausto Vagnetti}
\affil{Dipartimento di Fisica, Universit\`a di Roma ``Tor Vergata'',\\
Via della Ricerca Scientifica 1, I-00133 Roma (Italy)}
\email{fausto.vagnetti@roma2.infn.it}

\author{Dario Trevese, and Roberto Nesci}
\affil{Dipartimento di Fisica, Universit\`a di Roma ``La Sapienza'',\\
Piazzale A. Moro 2, I-00185 Roma (Italy)}
\email{dario.trevese@roma1.infn.it, roberto.nesci@uniroma1.it}

\begin{abstract}
Light curves  of eight BL Lac objects in the BVRI bands
have been analyzed. All of the objects tend to be bluer when brighter.
However spectral slope changes differ quantitatively from those of a sample of QSOs
analyzed in a previous paper \citep{tre02} and appear 
consistent with a different nature of the optical continuum.
A simple model representing the variability of a synchrotron component
can explain the spectral changes. Constraints on a possible thermal
accretion disk component contributing to the optical luminosity are discussed.
\end{abstract}

\keywords{galaxies: active - BL Lacertae objects: general -
BL Lacertae objects: individual (3C 66A, PKS 0422+004, S5
0716+71, OJ 287, ON 231, OQ 530, S5 1803+78, BL Lac) - 
quasars: general}

\section{INTRODUCTION}

Multi-wavelength observations of active galactic nuclei (AGN)
suggest the presence of various emission components with different relative
weights in different AGN classes. The large number of parameters involved in 
physical models of  emission mechanisms can hardly be constrained by
single epoch observations. Multi-band variability studies are then crucial in 
identifying and constraining individual components. Moreover, active nuclei are 
not spatially resolved in general. For this reason variability time scales,
delays and correlations between variations in different continuum components or 
emission lines provide the main  information about the location of 
different components.
The very definition of  AGN classes is in part based on variability properties.
In fact, while quasars (QSOs) and Seyfert galaxies vary, on average,  less than
0.5 mag on time scales of months in the optical band, blazars (i.e. BL Lac objects
and Optically Violent Variables, OVVs) may change their luminosities
by more than one mag in a few days \citep{ulr97}.
Quantifying spectral variability will then help understanding
different AGN classes by comparison.
Although monitoring campaigns, from radio to X-ray wavelengths, exist
for a limited number of objects, they are difficult to realize with adequate time 
sampling and duration. 
For this reason we try to exploit multi-band variability studies limited to  the optical
range, which in some case are available for several objects and relatively long observing 
periods.
In a previous paper \citet{tre02} analyzed the spectral slope variability
of a sample of PG QSOs on the basis of the light curves in the B and R bands,
provided by the Wise Observatory group, who monitored 42 quasars for 7 years
with a typical sampling interval of about one month \citep{giv99}. 
A spectral variability parameter $\beta \equiv \Delta \alpha / \Delta \log f_{\nu}$, 
$f_{\nu}$ being  the specific flux and $\alpha\equiv \partial \log f_{\nu} / \partial \log \nu$ 
the spectral slope, was defined by \citet{tre02} who derived constraints on variability mechanisms 
from the average $\alpha$ and $\beta$ values. In particular they showed that the variation of
spectral slope implied by changes of the accretion rate are not able to explain
$\beta$ values deduced from the observations, while ``hot spots'' on the accretion disc,
possibly caused by instability phenomena \citep{kaw98},
can easily account for the observed spectral variability.
In the present study we analyze B, V, R, I light curves of eight BL Lac objects 
obtained by \citet{dam02} in a monitoring campaign of about 5 years with an average
sampling interval of  $\approx 25$  days.
We show that  BL Lacs clearly differ from QSOs in their
$\alpha,\beta$ distribution. We  show that a simple model representing the variability 
of a synchrotron component can account for the observed $\alpha$ and $\beta$ values
and we discuss  constraints which can be derived from spectral variability
on the amount of thermal radiation contributed by a disk component in the optical band.

\section{DATA ANALYSIS}

The sample consists of 8 radio selected BL Lac objects listed in Table 1.
Observations were made in the period 1995-1999, in the standard Johnson-Cousins
B, V, R, I bands and are described in \citet{dam02}. Typical 
photometric uncertainties range from
0.01 to 0.03 mag r.m.s.. 
In table 1 the following quantities are reported:
{\it Column 1} - Name;
{\it Column 2} - Coordinate Designation; 
{\it Column 3} - Number of observations;
{\it Column 4} - average B magnitude;
{\it Column 5} - redshift;
{\it Column 6} - average spectral slope;
{\it Column 7} - spectral variability index.
For the present analysis we regard as ``simultaneous''  the observations in different bands
taken during the same observing night. We considered the data dereddened as in \citet{dam02}. 
We then define for each object an ``instantaneous'' slope
by a straight line fit through the B, V, R, I data points 
as a function of $\log \nu$,  neglecting  for the moment any curvature of the spectral
energy distribution within the optical range. 
A positive correlation of the instantaneous spectral slope   $\alpha\equiv \partial \log f_{\nu} / \partial \log \nu$
with flux changes  $\delta \log f_{\nu_R}$ has been already discussed by \citet{dam02}. 
In the present paper we quantify spectral variations in two different ways.
A first method considers, for each object,  all the positions in the $\alpha - \log f_{\nu_{B}}$ diagram,
$f_{\nu_{B}}$ is the specific flux at the effective
frequency $\nu_B$ of the blue band. 
These points, together
with the relevant regression lines  are shown in Figure 1 for each of the 8 objects.
In the same figure the solid lines represent the average increase of $\alpha$ with $\log f_{\nu_B}$.
Clearly this quantity is independent of the time at which  $\alpha$ and  $\log f_{\nu_B}$ values are measured.
We refer to the blue band, instead of the red one used by \citet{dam02}
in their correlation analysis, 
for comparison with the study of QSO spectral variations  by \citet{tre02}.
The slopes $b$ of the regression lines of each object can be assumed as a first spectral variability index.

A second method, adopted by \citet{tre02} considers, for each object,  the quantity:
\begin{equation}
\beta_{ij} \equiv \beta(\tau_{ij}) \equiv \frac{\alpha_j-\alpha_i}{\log f_{\nu_B}(t_j) - \log f_{\nu_B}(t_i)},
\qquad \tau_{ij} \equiv \frac{(t_j-t_i)}{(1+z)}, i,j=1,N
\end{equation}
which represents the  change of spectral slope per unit variation of log flux as a function 
of the rest frame time lag $\tau_{ij}$ and again we refer to the flux changes in the blue band.
We stress that $\beta$ is a very  sensitive indicator of spectral changes but, 
for this reason, it is also a noisy quantity.
For flux changes comparable or smaller than the photometric noise, 
$\beta$ values are randomly spread over a wide range. Thus we limit the computation to 
$| \log f_{\nu_B}(t_j) - \log f_{\nu_B}(t_i)| > 0.03$.
In any case, instead of individual $\beta_{ij}$ values, it's worth referring to average values,
e.g. in bins of time lag $\tau$.

To constrain physical models of variability, it's interesting to see whether
spectral changes  are different 
on different time scales. Figure 2 shows for each object
$\beta_{ij}$ as a function of $\tau_{ij}$.
The mean values of $\beta_{ij}$ in bins of 200 days are also shown.
The errors represent the standard deviation of the mean, which has been computed according to \cite{ede88},
to take into account correlation among $\beta_{ij}$ values within each bin.
For each object the value of the first spectral variability index $b$ is also shown for 
comparison.
In general the larger variations of $\beta$ occur at larger time lags where, due to the worst statistic, they are less
reliable. For $\tau \la 500$ d there are no evident systematic trends of $\beta$ as a function of $\tau$.
Thus it makes sense to define a less noisy index taking the average of $\beta$ within the first 500 days:
\begin{equation}
\beta \equiv \langle \beta_{ij} \rangle, ~~~~\tau_{ij} < 500 {\rm ~d}.
\end{equation}
In the following we limit our analysis to the average variability index $\beta$ .

Besides  the spectral slope $\alpha$ computed
by a straight line fit through the $\log f_{\nu_i}$ versus $\log \nu_i$  (i=B,V,R,I) data points, 
we also computed two other slope values,  towards  the blue and red side of the
spectral range:
$$\alpha_{BR}=\frac{0.4[(B-R)-(B_o-R_o)]}{\log ({\lambda_B}/{\lambda_R}) }-2, \quad
  \alpha_{VI}=\frac{0.4[(V-I)-(V_o-R_o)]}{\log ({\lambda_V}/{\lambda_I}) }-2,$$
where $B_o, V_o, R_o, I_o$ and $\lambda_B,\lambda_V,\lambda_R,\lambda_I$ are the zero points and the
effective wavelengths of the four photometric bands \citep{cox00}.
For all but two objects the values of $\alpha_{VI}$ and $\alpha_{BR}$ are respectively slightly larger and slightly
smaller, on average, than $\alpha$, 
indicating a little curvature of the spectrum, and the slopes of the linear regressions $\alpha_{BR}$ vs. $\alpha$ and $\alpha_{VI}$
vs. $\alpha$ do not differ significantly from one. 
In the case of BL Lac and OQ 530, $\alpha_{VI}$ is almost equal to $\alpha_{BR}$, or slightly smaller,  
and becomes  significantly smaller than $\alpha$ (and $\alpha_{BR}$) when $\alpha$ is small.
A possible explanation is that the (constant) contribution of the host galaxy emerges in the faint phases
causing a more marked softening of the spectrum on the red side of the wavelength interval considered.
This is consistent with the fact that the host galaxy contribution, as reported by \citet{urr00}, is relatively
strong for these two objects.

\subsection{Comparison with QSOs}

The average values $\beta$ of the  variability index, for all time lags smaller than 500 days rest-frame,
 are plotted in Figure 3 versus the relevant average $\alpha$ values. 
For comparison the values for the sample of 42 quasars 
discussed by \citet{tre02} are also shown\footnote{$\beta$ values for QSOs
are slightly different from those shown in \citet{tre02} since they have been recomputed at a fixed {\it rest-frame} time lag interval,
and taking the average over 500 instead of 1000 days for consistency.}.
BL Lacs and quasars are clearly segregated in the $\alpha-\beta$ plane. BL Lacs show, on average, 
lower $\alpha$ values, consistently with the absence, or lower relative 
weight, of the thermal  blue bump  component.
The new information in Figure 3 is that BL Lacs have $\beta$ values  
smaller, on average, than quasars. BL Lac variability is stronger, $\alpha$ changes are 
also larger, but their ratio $\beta$ is smaller. Moreover, using both $\alpha$ 
and $\beta$ parameters together allows a clear distinction of the two classes.
We want to discuss whether this behavior fits into the currently accepted models
describing the different emission mechanisms of blazars and QSOs.

\section{VARIABILITY OF THE SYNCHROTRON COMPONENT}

\subsection{A simple model}
The continuum spectral energy distribution of blazars from radio frequencies
to X and $\gamma$-rays can be explained by a synchrotron emission plus inverse Compton scattering
\citep{mar92,sik94}. Variability
can be produced by an intermittent channeling into the jet of the energy produced by 
the central engine.
\citet{spa01} have considered a detailed model where 
crossing of different shells of material, ejected with different velocities,
produce shocks which heat the electrons responsible for the synchrotron emission.
The resulting spectra are compared with multi-band, multi-epoch observations of
3C 279 from radio to $\gamma$ frequencies, showing a good agreement.
Similar computations have been done for Mkn 421 \citep{gue02}.
In the case of the eight objects of our sample,  B, V, R, I bands
are  sampling variability of the synchrotron component.
This component can be roughly described by a broken power law 
characterized by the break (or peak in $\nu L_{\nu}$) frequency $\nu_p$ and the asymptotic spectral slopes
$\alpha_1$ and $\alpha_2$ at low and high frequency respectively,
$\alpha_1 > -1$, $\alpha_2 < -1$  \citep{tav98}.
An equivalent representation is:
\begin{equation}
L_{\nu}=L_p \cdot 2[(\frac{\nu}{\nu_p})^{-\alpha_1}+(\frac{\nu}{\nu_p})^{-\alpha_2}]^{-1} \equiv L_p \cdot H(\nu/\nu_p;\alpha_1,\alpha_2)
\end{equation}
where $L_p$ is the specific luminosity at $\nu_p$.
We adopt this representation for a stationary component of the SED and we add to it
a second (variable) component with the same analytical form but different peak frequency $\nu_{p'}$  
and amplitude $L_{p'}$ to produce spectral changes:

\begin{equation}
L_{\nu}= L_p \cdot H(\nu/\nu_p; \alpha_1, \alpha_2) + L_{p'} \cdot H(\nu/\nu_{p'}; \alpha_1, \alpha_2)
\end{equation}

The addition of the second component mimics the behavior of the synchrotron emission
of model spectra when shell crossing occurs,  producing an increment of emission with
$\nu_{p'} > \nu_p$, due to newly accelerated electrons.
With such a representation we can compute $\alpha$ and $\beta$ as a function of 
$\nu_{p'}/\nu_p$, for different values of  $\nu_p$ and given values of
 $\alpha_1$, $\alpha_2$ and $L_{p'}/L_p$.  
We adopted typical values of the asymptotic slopes 
$\alpha_1=- 0.5$, $\alpha_2=-2$, we assigned to $\nu_p$ different values in the range
 2.5-10 $\cdot 10^{14}$ Hz, and we made  $\nu_{p'}/\nu_p$ vary in the range 0.8-4.
We also adopted  a ratio  $L_{p'}/L_p$ 
corresponding to a 0.5 mag change in the blue band, which is representative of the r.m.s. variability.
The results are shown in Figure 4, where  we use as $\alpha$ the slope of the 
stationary component. The three  different lines are computed for $\log \nu_p=14.4, 14.7, 15.0$
from left to right respectively. 
The computed curves fall naturally in the region occupied by the
data, namely it is possible to account for the position of the objects
in the $\alpha-\beta$ plane with typical values of  $\alpha_1$, $\alpha_2$, $\nu_{p'}/\nu_p$ and $\nu_p$,
required to represent the overall SED of the objects considered \citep[see][]{fos98}.

\subsection{The Thermal Bump}

BL Lac objects can be classified according to the break frequency of  
synchrotron emission, from Low frequency peaked  BL Lacs (LBL)  which
show a break in the IR/optical band to High frequency peaked  BL Lacs (HBL) 
with the synchrotron break at UV/X-ray (or higher) energies \citep[e.g.][]{pad95}.
On average the blue bump (TB) produced by the accretion disk
tends to be negligible in HBLs
while in LBLs its contribution may be important.
According to \citet{cav01} this can be interpreted in terms of higher
accretion rate $\dot m$ in the case of LBLs, producing a stronger
thermal emission from the disk, while HBLs are dominated by
non-thermal energy, partly  extracted from rotational energy of the central Kerr hole.

In particular the TB remains negligible during changes of the
synchrotron component in the case of Mkn 421, a classic HBL, while emerges
clearly during the faint phases in the case of 3C 279, a low frequency peaked object
\citep[see][]{spa01,gue02}. 
We stress that the TB is relatively ``narrow'',
since its emission is mostly concentrated in the UV region, while the synchrotron emission extends from
radio frequencies to the UV (or X-rays).
Then the change of relative intensity of the two components can produce strong
changes of the spectral slope in the optical-UV region, where the synchrotron and TB components
have very different slopes ($\approx -2$ and $\approx+1.5$ respectively). 
As a result, in the case of LBLs we could find strongly negative $\beta$ values in the faint 
phases, namely a hardening instead of a softening of the spectrum, at wavelengths
smaller but close to the peak of the TB,  as observed in the case of 3C 279, where we can estimate
$\beta \simeq -1$ at $\lambda \simeq 2500$ \AA~ on the basis of the data reported by
\citet{pia99} (see their Figure 4).
However these negative $\beta$ values do not seem to be typical in the case of our sample, although
synchrotron breaks  are distributed in the frequency range $10^{13} - 10^{15}$ Hz.
Thus, in principle,  we can set upper limits to  contribution of the accretion disk to the 
overall SED on the basis of the observed $\beta$ values. 
For a rough estimate it is sufficient to approximate the TB with a black body emission.
As already noted in \S 1, although the TB itself may vary, 
possibly due to instabilities which produce hot spots 
in the accretion disk \citep{tre02},
the amplitudes of its variations are smaller than those of blazars \citep[e.g][]{tre94}.
Thus we can approximately consider TB emission as a constant during the much larger variations
of the synchrotron component.

Thus we consider a specific luminosity $L_s \equiv L(\nu,t)$ of the synchrotron component
with a time dependent local slope $\alpha_{s}(t)$
at the observing frequency $\nu=\bar{\nu}$, and a constant thermal TB luminosity $L_{TB}$ with local slope $\alpha_{TB}$.
The total specific luminosity at two times $t$ and $t_o$ will be $L=L_s+L_{TB}$ and $L_o=L_{so}+L_{TB}$ respectively.
The local spectral slope $\alpha$ of the composite SED is:
\begin{equation}
\alpha \equiv \frac {\partial \log  L }{ \partial \log \nu} =
\frac {L_s}{L} \alpha_s + \frac {L_{TB}}{L} \alpha_{TB}
\end{equation}
Defining $\chi = \frac {L_{TB}}{L_{so}}$,  $\varepsilon = \frac {L_s}{L_{so}}$
and $\delta = \alpha_s - \alpha_{so}$,
where the subscript ``o'' indicates the initial time, we have $\beta_s = \delta / \log \varepsilon$
for the ``unperturbed'' synchrotron component and:
\begin{equation}
\beta\equiv  {(\alpha-\alpha_o)}/ {\log {\frac{L}{L_o}}}=
\left[ {\frac {\varepsilon}{\varepsilon+\chi}\delta+\frac {\chi(\varepsilon-1)}
{(\varepsilon+\chi)(1+\chi)}(\alpha_o-\alpha_{TB})}\right] \bigg{/} {\log \frac{\varepsilon+\chi}{1+\chi}}
\end{equation}
for the composite spectrum.
In Figure 5 we report the spectral variability index $\beta$ as a function of $\chi$
for different values of the temperature $T$.

We assume a typical value $\alpha_{so} =-1.75$ 
and  we assume for the disk a black body temperature $T= 10^4 K$ or  $T=2\times 10^4 K$.
In both cases we compute $\beta$ as a function of $\chi$, for two values of
$\varepsilon$ and a typical value $\beta_s=1$.
The result is shown in Figure 5. 
For 3C 279 we can estimate \citep[from][]{pia99}
$\beta \approx -1$, $\chi \approx 1$ and $T \approx 2\times 10^4 K$
which is consistent with the relevant curve in the figure.  
It appears that e.g. the condition $\beta > 0.5$ implies $\chi \la 0.4$ for $T= 10^4 K$,
or $\chi \la 0.2$ for $T=2\times 10^4 K$. Similarly $\beta > 0$ implies $\chi \la 1$ for $T= 10^4 K$,
or $\chi < 0.5$ for $T=2\times 10^4 K$.
For a given $\nu_{p}$, the ratio between TB and synchrotron components may, in fact, vary from typical BL Lacs
to flat spectrum radio quasars (FSRQ), which also show intense broad emission lines.
This different behavior depends both on the intrinsic luminosity of these components and
on the viewing angle. An estimate of the isotropic emission can be deduced from the 
broad line region luminosity $L_{BLR}$ \citep{cel97}
and compared with the kinetic jet power \citep{del03}, as done by \cite{pad03}.
This information is crucial in studying the relation
among different classes of objects and testing possible AGN ``grand unified theories''.
In this context, spectral variability indexes, possibly measured for stastistical samples
at different time lags, luminosities, and rest-frame frequencies, will provide independent
information on the relative importance of isotropic and beamed components.

\section{SUMMARY AND CONCLUSIONS}
We have performed a new analysis of the light curves of eight BL Lac objects
monitored in B,V,R,I bands for a total period of about 5  years with an average
sampling interval of $\approx 25$ days.

Adopting a spectral variability index $\beta$, representing the average ratio between the change of spectral slope and
the logarithmic luminosity change, we find that BL Lacs show smaller spectral variability with respect to
quasars (analyzed in a previous paper).
This happens despite the fact that both the spectral slope and luminosity variations of BL Lacs are larger then those of QSOs.
BL Lacs and QSOs appear clearly segregated in the $\alpha-\beta$ plane.
Our analysis allows a quantitative comparison of the observations with a model based on variability
of the synchrotron component. The model easily accounts for the observed position in the $\alpha-\beta$
plane with a natural choice of the relevant parameters. Thus the segregation in the $\alpha-\beta$ plane
is a consequence of the different emission mechanism in the optical band: synchrotron in the case of BL Lacs and
thermal hot spots on the accretion disk in the case of QSOs \citep[see][]{tre02}.
A strong thermal bump can produce negative $\beta$ values, which are not observed in our sample, while they do
in fact appear in some typical LBL object, where the thermal bump emerges in the faint phases.
Thus spectral variability, even restricted to the optical band, can be used to set limits on the relative
contribution of the synchrotron component and the thermal component to the overall SED.
Further studies of the spectral variability index $\beta$ in different objects will allow us to establish whether
the thermal bump is significant only in the most extreme LBL, or to what extent its contribution,
possibly correlated with the accretion rate $\dot m$, is related to total luminosity.

\acknowledgments
We wish to thank Alfonso Cavaliere and Enrico Massaro for discussions and comments, and
Silvia Sclavi for providing details on photometric calibrations.
This work was partly  supported by MIUR under grant COFIN 2001/028773.

{}

\clearpage 

\begin{figure}
\plotone{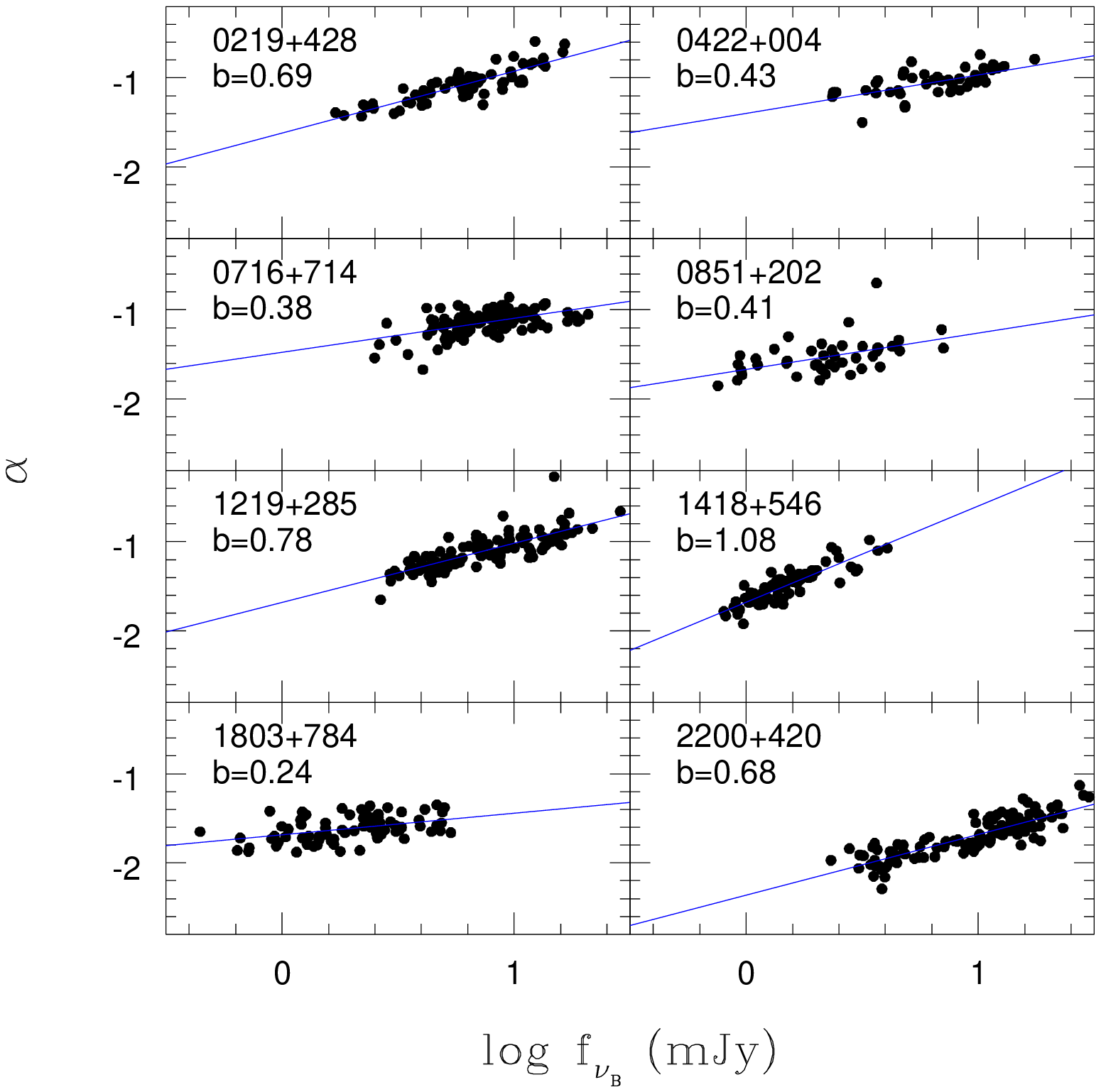}
\caption{Instantaneous spectral slope $\alpha$ versus specific flux at the  effective frequency 
of the observed blue band, for the  eight objects of the sample. The regression lines are also shown.
The slopes $b$, reported in the relevant panels, can be adopted as a first spectral variability indicator. \label{fig1}}
\end{figure}

\clearpage 
\begin{figure}
\plotone{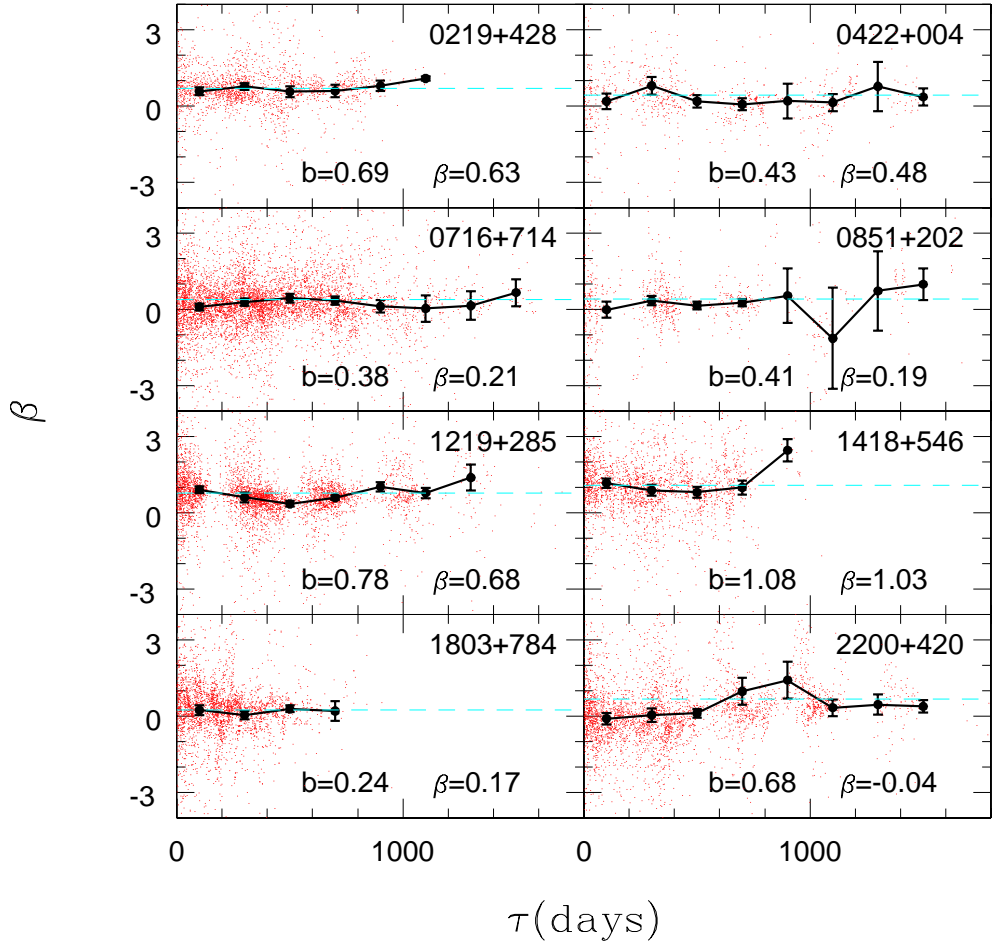}
\caption{$\beta_{ij}$ as a function of $\tau_{ij}$ for each object in the sample. Large dots represent
the mean values of $\beta_{ij}$ in bins of 200 days. Dashed lines correspond the first spectral variability indicator $b$. The values of $b$ and $\beta$ are also indicated.
\label{fig2}}
\end{figure}

\clearpage 
\begin{figure}
\plotone{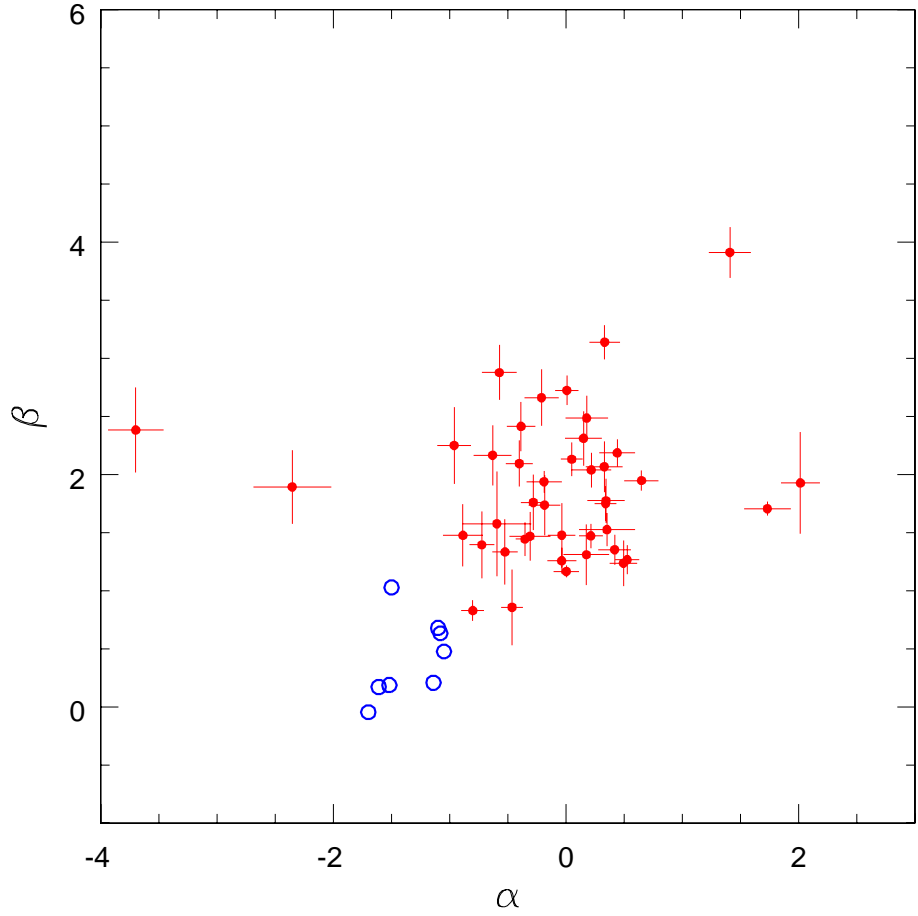}
\caption{The average values $\beta$ of the  variability index versus the
average spectral slope for the eight objects of the BL Lacs sample
({\it open circles}) and the 42 QSOs \citep[see][]{tre02} ({\it dots}).
\label{fig3}}
\end{figure}

\clearpage 
\begin{figure}
\plotone{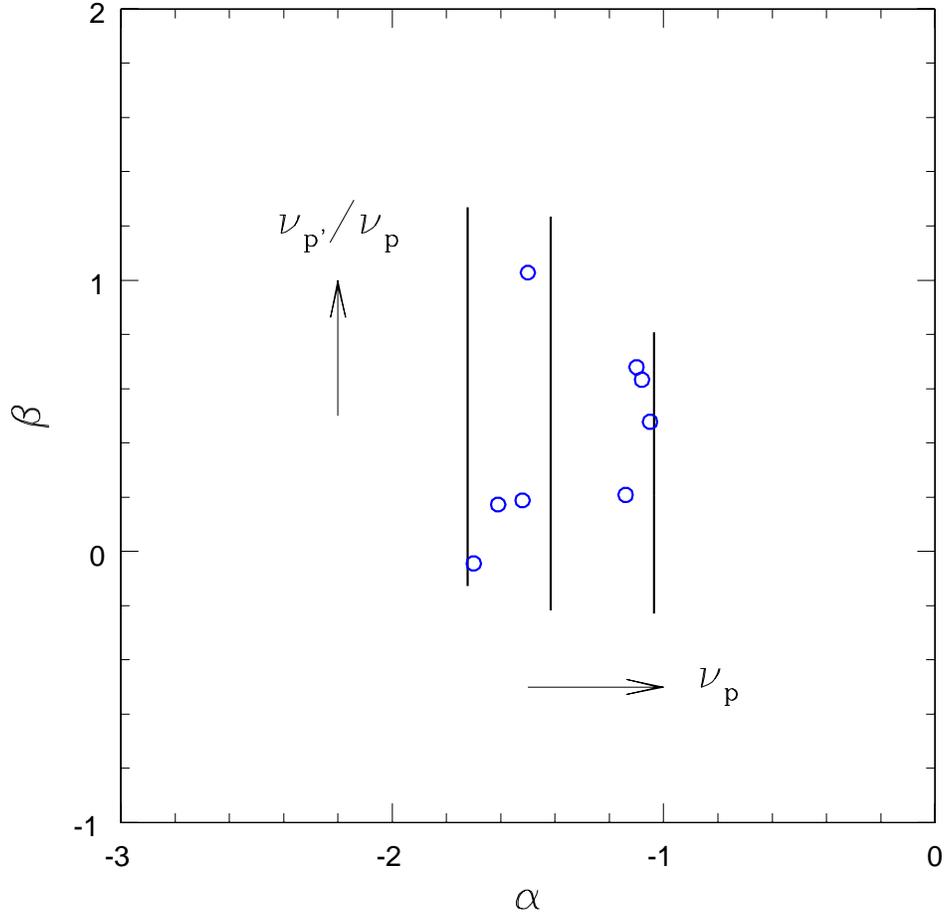}
\caption{Spectral variability index $\beta$ versus spectral slope. Open circles represent
BL Lac objects as in Figure 3. Lines represent the models. Each line corresponds to a fixed value
of $\nu_p$:  $\log \nu_p = 14.4, 14.7 , 15.0$.
Along each line the ratio $\nu_{p'}/\nu_p$ varies from 0.8 to 4 (from bottom to top).
We adopted an $L_{p'}/L_p$ ratio corresponding a magnitude change of 0.5 mag.\label{fig4}}
\end{figure}

\clearpage 
\begin{figure}
\plotone{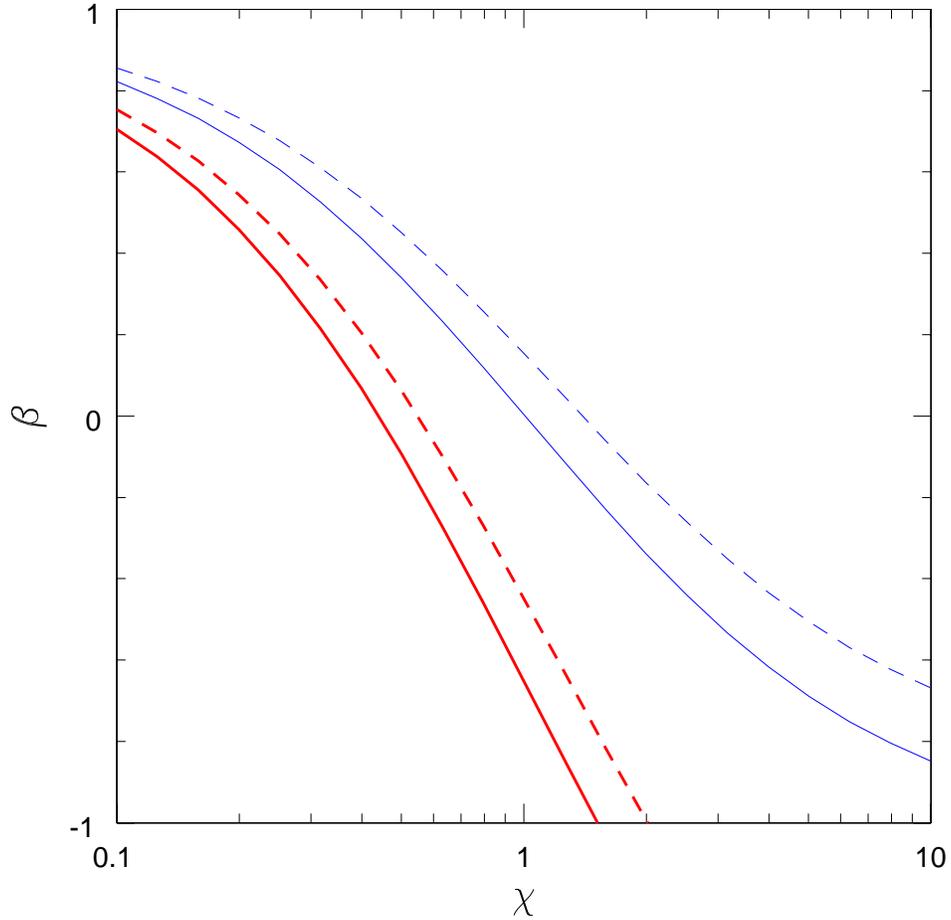}
\caption{
Spectral variability versus the ratio between the thermal bump and synchrotron 
components at the observing frequency: $\chi = \frac {L_{TB}}{L_{so}}$. The curves have been computed for 
two values of the thermal component temperature $T=2 \cdot 10^4$ K ({\it thick lines}) and $T= 10^4$ K
({\it thin lines}), and for $\varepsilon = 1.1$ ({\it continuous lines}) $\varepsilon = 1.5$ 
({\it dashed lines}). \label{fig5}}
\end{figure}

\clearpage

\begin{deluxetable}{lcccccc}
\tabletypesize{\scriptsize}
\tablecaption{The sample of BL Lac objects. \label{tbl-1}}
\tablewidth{0pt}
\tablehead{
\colhead{Name} & \colhead{Coord. Desig.}   & \colhead{$N_{obs}$}   &
\colhead{B}   &  \colhead{z} &
\colhead{$\alpha$}  & \colhead{$\beta$}  
}

\startdata
3C 66A		& 0219+428 &~66&  15.01	& 0.444 & -1.08 & ~0.63 \\
PKS 0422+004	& 0422+004 &~44&  15.00	& -     & -1.05 & ~0.48 \\
S5 0716+71	& 0716+714 &118&  14.51 & -     & -1.14 & ~0.21 \\
OJ 287		& 0851+202 &~44&  15.83 & 0.306 & -1.52 &  0.19 \\
ON 231		& 1219+285 &101&  14.50 & 0.102 & -1.10 & ~0.68 \\
OQ 530		& 1418+546 &~74&  16.22 & 0.152 & -1.50 & ~1.03 \\
S5 1803+78	& 1803+784 &~84&  16.07 & 0.680 & -1.61 & ~0.17 \\
BL Lac		& 2200+420 &120&  15.56 & 0.069 & -1.70 & -0.04 \\
\enddata
\end{deluxetable}
\end{document}